# Auxiliary Optomechanical Tools for 3D Cell Manipulation

**Ivan Shishkin[1],\***, **Hen Markovich[2,4]**, **Yael Roichman[3,4]**, **Pavel Ginzburg** [2,4]

[1] ITMO University, Saint-Petersburg, 197101, Russia
[2] School of Electrical Engineering, Tel Aviv University, Tel Aviv, 69978, Israel
[3] School of Chemistry, Tel Aviv University, Tel Aviv, 69978, Israel
[4] Light-Matter Interaction Centre, Tel Aviv University, Tel Aviv, 69978, Israel
\* Correspondence: i.shishkin@metalab.ifmo.ru

**Abstract:** Advances in laser and optoelectronic technologies brought the general concept of optomechanical manipulation to the level of standard biophysical tools, paving ways towards controlled experiments and measurements of tiny mechanical forces. Recent developments in direct laser writing (DLW), enabled the realization of new types of micron-scale optomechanical tools, capable of performing designated functions. Here we further develop the concept of DLW-fabricated optomechanically-driven tools and demonstrate full-3D manipulation capabilities over biological objects. In particular, we resolved a long-standing problem of out-of-plane rotation in a pure liquid, which was demonstrated on a living cell, clamped between a pair of forks, designed for efficient manipulation with holographic optical tweezers. The demonstrated concept paves new ways towards realization of flexible tools for performing on-demand functions over biological objects, such as cell tomography and surgery to name just few.

## 1. Introduction

Three-dimensional optical microscopy techniques are invaluable tools in modern biomedical and biophysical studies. Imaging techniques include, for example confocal [1], multiphoton[2] and super-resolution[3] microscopy. In most cases, these methods require immobilization of objects under study, at least during image acquisition. The common way to obtain three-dimensional (3D) characterization of the investigated objects is to combine fluorescent dyes and image sectioning. This requires additional sample preparation and in some cases may harm or alter the studied system. In biomedical diagnostics, it is especially beneficial to obtain such characterization in a label free manner. For this reason various 3D tomography techniques, based on quantitative phase microscopy, were developed in the past few years. For example, single-cell optical coherence tomography can be implemented [4], however, imaging-based techniques appear more practical. These techniques require the ability to rotate an object in suspension in a controlled manner. Sample scanning can be achieved by several methods, including loading objects under study in gel-filled microcapillaries and rotating them mechanically[5,6]. Optical tweezers also have been demonstrated as a viable tool for object scanning by using single-beam time-shared trap[7] (suitable only for elongated objects like E. coli), and by using multiple optical traps applied to non-spherical objects like diatoms [8] and yeast cells [9].

The possibility of the rotation of individual cell was demonstrated using a pair of counterpropagating beams from single-mode fibers inserted in the microfluidic channel [10–12], using holographic optical trap [13] or electrorotation[14]. However, it should be noted that direct or trapping of living tissue with counterpropagating laser beams or with structured light in holographic tweezers is constrained by localized heating [15] and phototoxicity [16,17].

Our approach provides capabilities of full 3D manipulation of biological samples within solutions and allows achieving a set of essential functionalities, including (i) prevention of photoinduced damage to living cells, (ii) manipulation of transparent/low contrast objects, (iii) 3D manipulation, including rotation, of spherical species. Furthermore, from the fundamental standpoint, we demonstrate the utilization of radiation pressure forces for achieving controllable rotation of objects. Our general concept is depicted in Fig. 1 (a), where optomechanically driven 'cell clamps' immobilize a biological cell. Those micron-size clamps are fabricated with the help of direct laser writing [18] (DLW). This technique is based on two-photon absorption [19] in photopolymerizable materials [20], which is an extremely viable method for fabrication of structures with sub-micron-scale resolution. A few notable examples of DLW-based structures for opto-fluidic applications include force and topography-sensing optically driven scanning microprobes [21–23], light-actuated microsyringes [24] and platforms for targeted light delivery to microscopic objects like cells [25]. It should be noted, that several designs were developed earlier to achieve out-of-plane rotation, like paddlewheel [26] and crankshaft-like structure [27], however none of them were tested yet for manipulation of individual cells.

Our auxiliary structures are driven into motion with the help of holographic optical tweezes. Each clamp is illuminated with three beams – a pair for immobilization and the third one for achieving the rotation of the trapped cell (revolver geometry). Those micro-tools allow clamping of an object, translating it towards the analyzing apparatus, rotating it, and finally releasing it back to the suspension. Furthermore, the immobilized cell is not directly illuminated by intense laser light and the whole scheme does not rely on cell's parameters, which makes this approach quite universal. We report on the design, fabrication and use of 3D printed unique cell clamps, that enable trapping, translation, and rotation of cells using optical forces focused away from the cell.

## 2. Materials and Methods

The proposed approach towards cell rotation is schematically depicted in Fig. 1(a). A pair of auxiliary tools fabricated by DLW is detached from substrate and immobilized with trapping laser beams. Afterwards the desired cell is located and is immobilized by a pair of tools driven into its proximity simultaneously.

A scanning electron microscopy (SEM) image of the designed cell clamps, shaped like forks, is presented in Fig. 1 (b,c). The structure has several essential elements – (i) a fork end for clamping a cell. (ii) Base and top spheres for optical trapping with gradient forces, these are used to control their position and orientation. The radius of the sphere is large enough compared to the fork core to ensure localized trapping [23]. (iii) Three spheres, forming a revolver operating like a windmill, are placed in between these two spheres. The distance between the centers of 'actuating' spheres and the axis of the symmetry of the structure was set 5 µm. The clamps are 30 µm long and each spherical feature is 5 µm in diameter. The distance between the base and top spheres used for immobilization of the tool was chosen to be 18 µm. It should be noted, that the dimensions of the microtool could be reduced roughly by a factor of 2, however larger size was chosen for better mechanical stability and to allow easier detachment with a micromanipulator. Before performing optical experiments, the coverslips with the microforks were cured overnight with a UV lamp in order to suppress residual fluorescence from the photoinitiator and to increase their mechanical stiffness.

Projection of a defocused trap on the 'revolver' part of the structure is the optimal way to use radiation pressure to rotate the fork around its axis. Rotation can be initiated by turning the third optical trap on, stopped by turning it off, and reversed by projecting the trap of the other side of the fork, in real time using a holographic optical tweezers (HOTs) setup [28]. The setup used green 532 nm laser, a reflective spatial light modulator (SLM) module, beam expanders and inverted bright field microscope. The laser beam was expanded in order to overfill SLM aperture. The SLM is placed in the focus of shrinking telescope, which forms 4f-system with the microscope objective. The zero order spot was blocked in the focal plane of the negative beam expander after the SLM. The beam is reflected upward inside the microscope using a beam splitter cube and is focused using a 100X Olympus oil immersion objective (NA = 1.4) into the sample chamber. The phase masks of the SLM for the trapping and manipulating (open, close and rotate) of the micro-tools were designed using MATLAB and calculated using Gerchberg–Saxton iterative algorithm. The traps placement has been designed to be symmetric with respect to the zero-order beam to reduce the intensity of the higher diffraction orders of the SLM.

Phosphate buffer saline (PBS) with 0.5% TWEEN 20 was used as working medium. The structures were mechanically detached from the coverslip using a glass microneedle connected to the micromanipulator (Scientifica Patchstar) before conducting the experiments. After detaching two structures from the surface, the shutter of the laser was opened. By using the motorized stage the structures are brought into the trapping focal spots of the laser and are immobilized. Since the access for the glass capillary was needed, the samples were not sealed in double-glass chamber and occasional addition of water was needed in order to compensate for the evaporation.

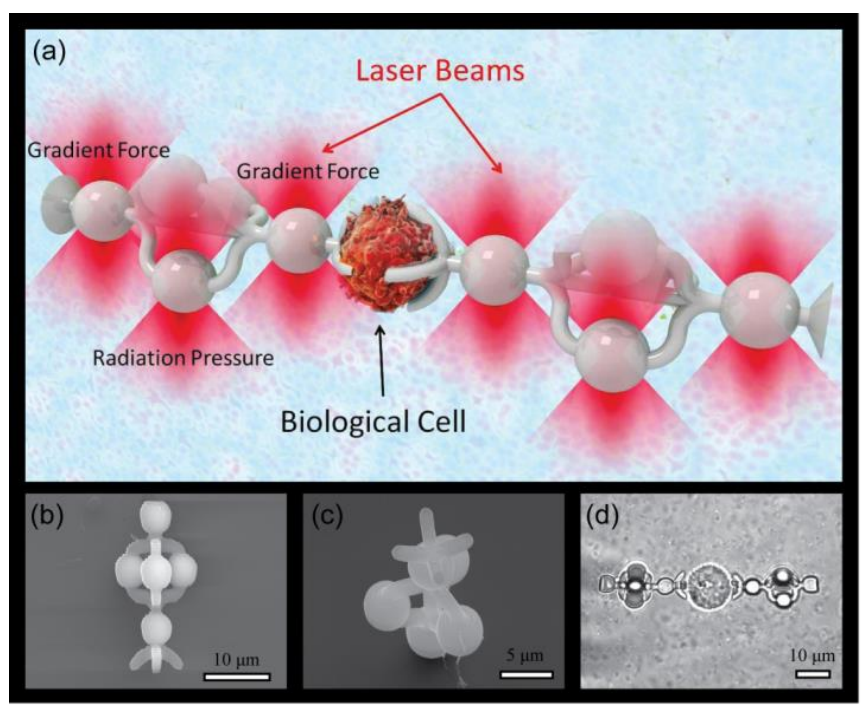

**Figure 1.** (a) An artist's view of the cell clamps at work. The fork-like shaped clamps are optically trapped in 3D by holographic optical tweezers. SEM side view (b) and top view (c) of fabricated auxiliary microtools. (d) Microscope image of microtools engaged in immobilization of sw480 adenocarcinoma cell.

**3. Results**

In order to demonstrate feasibility of the proposed approach towards object manipulation, it was necessary to show the capability of rotation of the individual micro-tool. After detachment from the coverslip, the tool was immobilized using two generated traps, positioned at anchoring points marked with red crosses in 1st frame in Fig. 2. After successful immobilization, the third 'actuator' trap was generated 3 microns off plane in the spot marked with the cross. The stable trapping of individual tool was achieved with laser power of 0.8W incident on SLM, which was distributed between three trapping spots. The relative power of the 'actuator' trap was reduced compared to the two main immobilization traps by 50% to improve stability of trapping.

The video sequence of the experiment with a single micro-tool is presented in Fig. 2 as set of individual frames captured with 1.5 second interval (see Visualization 1 for complete sequence). It might be clearly seen, that with the proposed configuration of the traps the desired axial rotation can be achieved. The rotational motion was induced with the radiation pressure force that pushes one of the beads in the revolver part. The photon momentum is transformed to the structure owing to light absorption in the polymer, which arises from residual molecules of photoinitiator and intrinsic material absorption. It is worth noting that translational motion in the plane of view of the trapping objective is can be demonstrated straightforwardly and is not shown here.

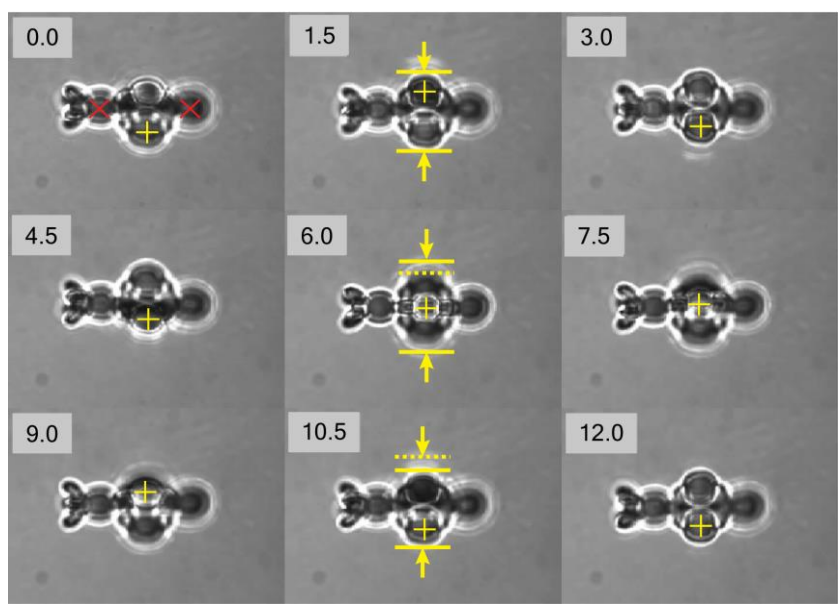

**Figure 2.** Frame sequence obtained from captured video demonstrating rotation of a trapped single micro-tool. Frames were extracted every 1.5 seconds. Red crosses in frame at 0 second mark immobilization trap positions. Yellow crosses mark positions of displaced microspheres. The solid and dotted yellow lines shown in frames for 1.5, 6 and 10.5 seconds reveal variance of one of the dimensions of projection of the microtool captured by camera.

We analyze the rotation dynamics and trapping stability of individual micro-tool by image analysis algorithms. Sufficient contrast between the microtool and the background allows to implement edge detection algorithm. Each captured frame of the recorded video is processed as follows – the edges of the object are detected using Sobel operator followed by dilation and filling of gaps in resulting image, which allows to obtain binary mask corresponding to the object. The properties of resulting binary images were analyzed using Matlab *regionprops* function– the center-of-mass (CoM) positions were extracted and axes of the equivalent ellipse with same normalized second-moments as of the original object binary image were obtained. These parameters were further used for analysis of motion of a single microtool.

The information on CoM position over each frame is presented as probability distribution in Fig. 3(a). The respective data on X and Y position distributions is presented in Fig.3(b). The stiffness of trapping potential was analyzed using equipartition theorem and allowed to obtain values of $k_x$ =2.22pN/µm and $k_y$ = 1.96pN/µm. The values of the trapping potential stiffness could be used for assessment of the maximum possible force which could be exerted on the immobilized cell in order to assess individual cell stiffness [29]

The equivalent ellipse minor axis variance over time is presented in Fig.3(c). The periodic variations of the value can clearly be observed, which can be attributed to rotation of the microtool. The corresponding Fourier spectrum of the extracted signal is presented in Fig.3(d). Fourier analysis of the time-varying parameter allowed to reveal processes with different periodicity – $T_1$ = 8.6 seconds (0.117 Hz), which corresponds to complete 360 degree revolution of the microtool and $T_2$ = 3.3 s (0.3Hz) which could be attributed to out-of-plane rotation of the tool by 120°.

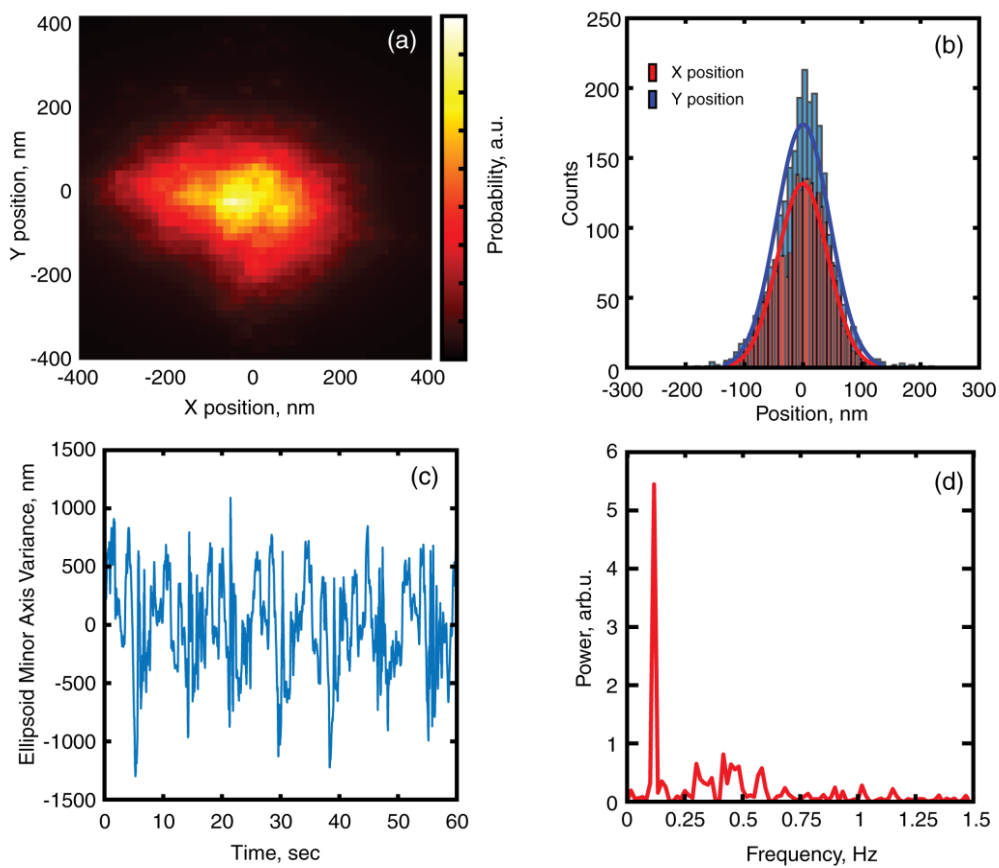

**Figure 3**. (a) Probability distribution of a center of mass (CoM) for single microtool. (b) Position distribution for X and Y coordinates of CoM. (c) Variance of equivalent ellipsoid minor axis, as a function of time (d) Power spectrum of the ellipsoid minor axis variance.

In order to demonstrate proof-of-concept capability of axial rotation of the living biological object, we have undertaken the experiments using yeast cells as the test object. For such experiment a pair of micro-tools were detached and immobilized in optical traps with relative separation of 15 microns between edges. After finding the object for studies, the tools were driven together to proximity (video is presented in Visualization 2), resulting in immobilization of the tested object. The success of the immobilization was checked by scanning with the microscope stage. The 'actuator' traps were turned on, resulting in simultaneous rotation of the trapped micro-tools, which transferred the torque on the object clamped between them. The frames of captured video sequence are presented in Fig. 4 (cell immobilization is shown in Visualization 3, cell rotation is shown in Visualization 4).

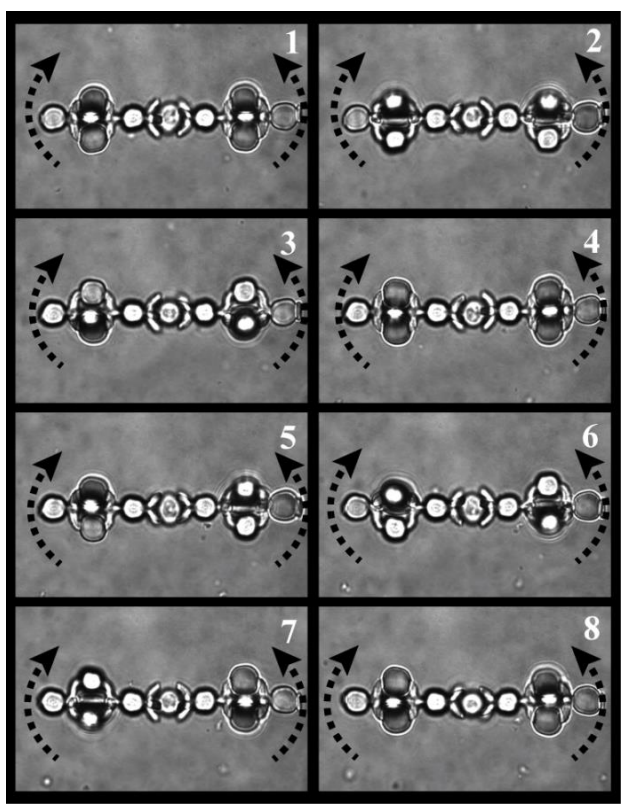

**Figure 4.** Frame sequence of rotation of trapped yeast cell. Frames were captured each second.

## 4. Discussion

The recorded videos and their frame-by-frame sequences demonstrate feasibility of the proposed approach auxiliary structures for micro-manipulation of objects. The analysis of motion of single microtool revealed the stability of microtool immobilization and out-of-plane rotation with the speed of 6-7 revolutions per minute was shown. The proof-of-principle manipulation of individual live cell with a pair of microtools was demonstrated as well. However, several problems should be addressed in order to successfully implement these microtools for more complex studies.

First, for the optical tomography applications, the rotation angle of the sample should be known. This can be achieved by synchronization of rotation of the auxiliary tools. For this, one needs to project holograms with out-of-plane position of 'actuator' trap in order to control the angle of revolution of individual microtool. The current implementation of trapping algorithm (e.g. static projection of 'actuator' trap) does not provide sufficient control over rotation speed of the individual tool and does not synchronize the motion of a pair of tools.

Second, the Brownian motion of tools in liquid results in drifts of the studied object not only in the XY-plane, but in Z-plane as well. The simple image processing technique implemented in this work allowed to determine centroid position of single microtool and can be further extended towards analysis of the motion of a pair of microtools with the cell immobilized in between them.

## 5. Conclusions

For the summary, we presented a new approach that will allow a complete 360 degree scan of the biological object in-vitro embedded in its host fluid environment. For example, optical diffraction tomography (ODT) [30] allows measuring the refractive index distribution of optically transparent object, such as cancer cells (our proof of concept result appears in Fig. 1(d)). The method does not require labeling or high intensity light sources. Crucially, the resolution of ODT depends on the range of angle from which imaging takes place. Two axis full rotation will be optimal. Two principal approaches exist towards the scan acquisition in ODT - illumination scanning [31] and sample scanning [32]. It should be noted that illumination scanning methods are constrained by limited projection angles [33]. Our proposed technique provides the possibility to do such two axis rotation by re-trapping a cell after one axis rotation is performed. Experiments combining our technique and ODT are underway. These developments can open new horizons in microscopy, where accurate full-three-dimensional mapping of biological objects and even other valuable functions can be performed with auxiliary optomechanically driven micro-tools.

**Acknowledgments:**

This research was funded by ERC StG 'In Motion' (802279).